\documentclass[apjl]{emulateapj}
\usepackage{subfigure}

\def\ltsim{~\rlap{\lower -0.5ex\hbox{$<$}}{\lower 0.5ex\hbox{$\sim\,$}}}

\slugcomment{To appear in ApJL}

\shorttitle{Nearby superdense massive galaxies}
\shortauthors{Trujillo et al.}

\begin{document}


\title{Superdense massive galaxies in the Nearby Universe}


\author{Ignacio Trujillo\altaffilmark{1}, A. Javier Cenarro, Adriana de
Lorenzo-C\'aceres, Alexandre Vazdekis, \\
Ignacio G. de la Rosa, Antonio Cava}
\affil{Instituto de Astrof\'isica de Canarias, E-38205, La Laguna, Tenerife, Spain}
\email{trujillo@iac.es}


\altaffiltext{1}{Ram\'on y Cajal Fellow}


\begin{abstract}

Superdense massive galaxies (r$_e$$\sim$1 kpc; M$\sim$10$^{11}$ M$_\sun$) were common in
the early universe (z$\gtrsim$1.5). Within some hierarchical merging scenarios, a
non-negligible fraction (1-10\%) of these galaxies is expected to survive since that epoch
retaining their compactness and presenting old stellar populations in the present universe.
Using the NYU Value-Added Galaxy Catalog from the SDSS Data Release 6  we find only a tiny
fraction of galaxies ($\sim$0.03 \%) with r$_e$$\lesssim$1.5 kpc and
M$_\star$$\gtrsim$8$\times$10$^{10}$ M$_\sun$ in the local Universe (z$<$0.2).
Surprinsingly, they are relatively young ($\sim$2 Gyr) and metal--rich ([Z/H]$\sim 0.2$).
The consequences of these findings within the current two competing size evolution
scenarios for the most massive galaxies ("dry" mergers vs "puffing up" due to quasar
activity) are discussed.

\end{abstract}




\section{Introduction}
\label{intro}

The discovery (Daddi et al. 2005; Trujillo et al. 2006) that the most massive
galaxies (M$_\star$$\gtrsim$10$^{11}$$M_{\sun}$), independently of their star
formation rate (P\'erez-Gonz\'alez et al. 2008), were much more compact in the
past (a factor of $\sim$4 at z$\gtrsim$1.5 than their equally massive local
counteparts; Trujillo et al. 2007, Longhetti et al. 2007, Zirm et al. 2007, Toft
et al. 2007, Cimatti et al. 2008, van Dokkum et al. 2008, Buitrago et al. 2008;
van der Wel et al. 2008)  forces us to face the question of how these high--z
galaxies have evolved into the present-day massive population. Most of the current
theoretical work in this area suggests  "dry" mergers as the dominant mechanism
for the size and stellar mass growth  of these very dense galaxies (Khochfar \&
Silk 2006; Hopkins et al. 2008). As cosmic time evolves, the high-z compact
galaxies are thought to evolve into the present-day cores of the brightest cluster
galaxies. However, a fraction of these objects might survive intact from early
formation having stellar population with old ages. In fact, in some model
renditions the fraction of massive objects that could last without having any
significant transformation since z$\gtrsim$2 could reach 1-10\% (with a space
density of $\sim$10$^{-4}$ Mpc$^{-3}$; Hopkins et al. 2008). Recently, however,
Fan et al. (2008) have suggested an alternative scenario where the size evolution
is related to the quasar feedback instead of merging. In this model, galaxies puff
up after loosing huge amounts of cold gas due to the quasar activity.

If the relic superdense massive galaxies exist in the nearby Universe (z$\ltsim$0.2) we
should be able to find several thousands in the Sloan Digital Sky Survey Data Release 6
(SDSS DR6) spectroscopy survey which covers 6750 deg$^2$ (or a total volume of
3.73$\times$10$^8$ Mpc$^3$  up to z=0.2). With this aim we have used the NYU Value-Added
Galaxy Catalog (Blanton et al. 2005a) to probe whether there is any nearby massive
(M$_\star$$\sim$10$^{11}$$M_{\sun}$) galaxy as compact (r$_e$$\sim$1 kpc)  as those found
at high-z, and if so,   study  their stellar populations to provide unique clues for
understanding the nature and evolution of these objects. In what follows, we adopt a
cosmology of $\Omega_m$=0.3, $\Omega_\Lambda$=0.7 and H$_0$=70 km s$^{-1}$ Mpc$^{-1}$.

\section{The Data}
\label{data}

Our sample was selected using the NYU Value-Added Galaxy Catalog
(DR6)\footnote{http://sdss.physics.nyu.edu/vagc-dr6/vagc0/}. This catalog include
photometric information for a total of $\sim$2.65$\times$10$^6$ nearby (mostly below
z$\sim$0.3) objects. Around 1.1$\times$10$^6$ of these objects have spectroscopic redshift
determination. In addition, the catalog contains information about effective radii (Blanton
et al. 2005b) and stellar masses (Blanton \& Roweis 2007) based on a Chabrier (2003)
initial mass function.

\begin{figure}[t]

\plottwo{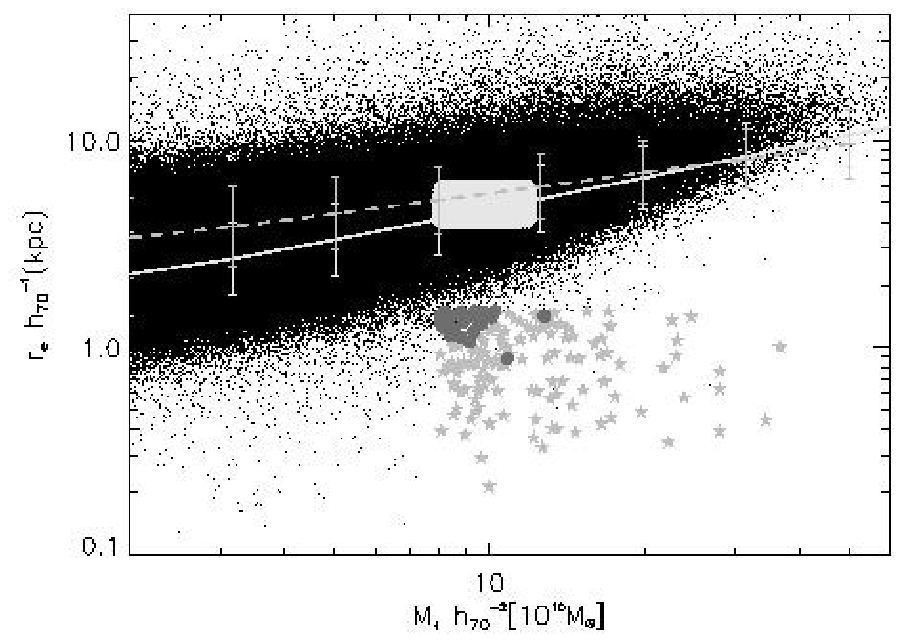}{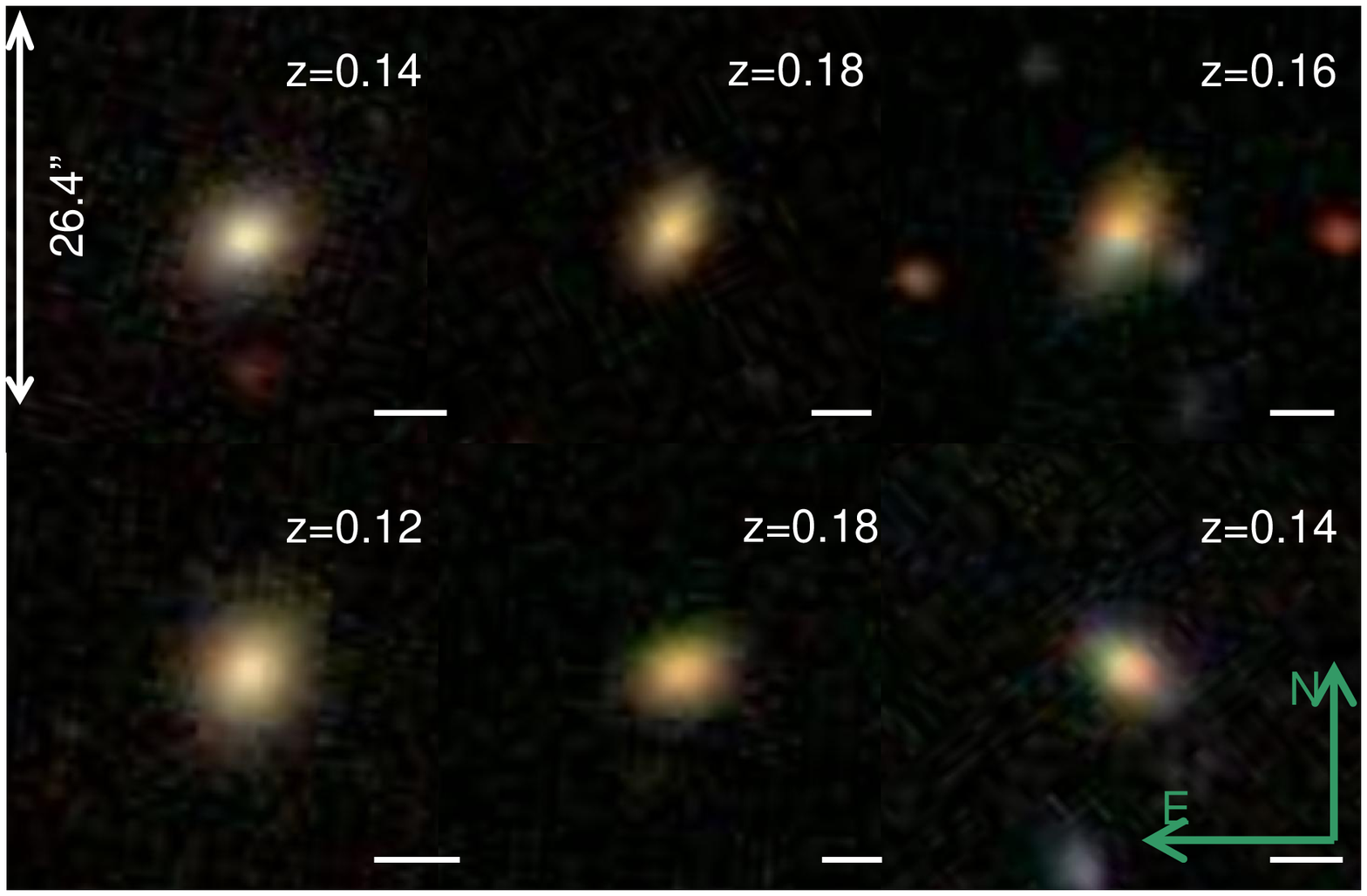}
\vspace{7cm}

\caption{\textit{Left}: Stellar mass--size distribution of the NYU Value-Added Galaxy
Catalog (DR6) galaxies. The position of our sample of compact galaxies is shown with 
circles. The stars are sources that according to the SDSS spectra classification are
QS0s. Following Shen et al. (2003), over-plotted on the observed distribution are the mean
and dispersion of the distribution of the S\'ersic half-light radius of the SDSS early-type
(n$>$2.5; solid line) and late-type (n$<$2.5; dashed line) galaxies as a function of the
stellar mass. The gray rectangular area shows the region used to extract the control
sample galaxies. \textit{Right}: Mosaic showing 6 typical galaxies from our sample. The
solid lines indicate the equivalent to 10 kpc size at the distance of the objects.}

\label{stelmass}
\end{figure}

With 0$<$z$<$0.2 and M$_\star$$>$8$\times$10$^{10}$$M_{\sun}$ there are 152083
targets. From these, 253 objects ($\sim$0.17\% of the above subsample) have
r$_e$$<$1.5 kpc according to the catalog. 56 of these objects are likely to be
point sources (i.e. stars) since their sizes estimates in the r-band are smaller
than 0.05 arcsec and are not considered in what follows.  This left a total number
of 197 targets ($\sim$0.13\%). As expected because of their compact nature, the
vast majority of these objects are QSOs (146) with a tiny contamination of stars
(3) according to the SDSS classification spectra, and are not considered in our
analysis since their sizes are not representative of their host galaxies. After a
detailed visual inspection of the remaining set of objects (48; i.e. only a 0.03\%
of our initial subsample) we find that 3 are very close to very bright stars, 8
are in close pair with other galaxies and 8 are edge-on disk galaxies where dust
effects can be relevant, preventing us from using them for a further analysis. 
Consequently, we remain with a final selection of 29 galaxies (see Table 1). The
mean redshift of these galaxies is $\sim$0.16 with a r.m.s. of 0.02. These objects
have a mean effective radius of $\sim$1.3 kpc (r.m.s. 0.15 kpc) and a mean stellar
mass of $\sim$9.2$\times$10$^{10}$$M_{\sun}$ (r.m.s.
1.2$\times$10$^{10}$$M_{\sun}$). Assuming a spherical symmetry this implies a
stellar density of $\sim$5$\times$10$^{9}$$M_{\sun}$kpc$^{-3}$ (this is just a
factor of $\sim$2 smaller than some globular clusters) and a stellar surface mass
density of $\sigma_{50}$$\sim$9$\times$10$^{9}$$M_{\sun}$kpc$^{-2}$. Fig. \ref{stelmass}
illustrates the position of the selected compact galaxies in the stellar mass-size
plane, as well as some examples of the galaxies in our sample.

\subsection{Robustness of stellar mass and size estimates}

To address the robustness of the stellar mass and the sizes of the 29 compact
galaxies of our sample, we have remeasured those quantities using different codes
than those employed in the NYU Value-Added Galaxy Catalog. To check the sizes we
used GALFIT (Peng et al. 2002). To calculate r$_e$, we used a S\'ersic 2D model
convolved with the PSF of the image (obtained from a nearby star to the source).
After circularizing the GALFIT sizes we find a good agreement between both
estimates, being the GALFIT measurement marginally smaller:
$\Delta$r$_e$/r$_e$=-0.17$\pm$0.14. The S\'ersic indices are also very similar
using both codes: $\Delta$n/n=-0.04$\pm$0.21, and with $<$n$>$$\sim$4.7. In
addition, to check whether our galaxies can be affected by a potential bias due to
a recent central burst which resulted in a understimation of our sizes, we have
compared the sizes found in the r-band against the sizes measured in the other
SDSS filters (u,g,i and z). In all the bands the galaxies show very similar sizes.
If anything, the galaxies are slightly larger in the u-band (i.e. against the
hypothesis of a central starburst):
$<$(r$_{e,r}$-r$_{e,u})$/r$_{e,r}$$>$=-0.19$\pm$0.07, although the result is not
statistically significant.  

The stellar masses were remeasured using the Bell et al. (2003) prescription based on the
restframe (g-r) color assuming a Kroupa IMF. We obtain a very good agreement with the NYU
Value-Added Galaxy Catalog estimates: $\Delta$M$_\star$/M$_\star$=0.02$\pm$0.08. In
addition to the stellar masses, we have compiled the central velocity dispersions of these
objects provided by the SDSS archive. We get a median central velocity dispersion of 196
km/s with a standard deviation of 30 km/s. We have also checked these values using the
Penalized Pixel-Fitting method developed by Cappellari \& Emsellem (2004) and we find an
excellent agreement with the SDSS measurements $\Delta$$\sigma$/$\sigma$=0.00$\pm$0.07.

\subsection{Control Sample}

In order to make a consistent analysis of the stellar population properties of the
compact galaxies we have created, using the same catalog, a control sample of
galaxies with similar stellar masses but with sizes representative of the average
galaxy population. These galaxies were selected using the following criteria:
0.8$<$M$_\star$$<$1.2$\times$10$^{11}$$M_{\sun}$, 4$<$r$_e$$<$6 kpc  and within a
volume of radius 30 Mpc in relation to the compact galaxies (to assure the
environmental conditions are similar). This produces a control sample of 299
objects after rejection of galaxies closer to bright stars and undergoing mergers.
The control sample shows a median central velocity dispersion of 180 km/s with a
standard deviation of 34 km/s, and a S\'ersic index $<$n$>$$\sim$3.8. Interestingly, the mean central velocity dispersion
of the control sample is slightly smaller than the mean central velocity
dispersion of the compact sample. This trend towards larger central velocity
dispersion, at a given fixed mass (luminosity), for those galaxies with smaller
sizes is also found in Bernardi et al. (2006).

\subsection{SDSS spectra}

The stellar population properties of our sample were analyzed using the spectra available in
the SDSS archive. The SDSS spectra cover 3800$<$$\lambda$$<$9200 $\AA$ and the fiber has
3$\arcsec$ in diameter. This implies that for a typical compact object of our sample we cover
the inner 3r$_e$ with a S/N of $\sim$30. A stacked spectrum of the galaxies in our sample is
shown in Fig. \ref{spectra}. We also show the stacked spectrum of the galaxies in our control
sample. For a typical galaxy in this control sample, the fiber covers the inner 1r$_e$ with a
S/N$\sim$20.

\subsection{AGNs}

A potential source of error in the size determination of any galaxy is
the presence of an active galactic nucleus (AGN) in its center which
can bias our measurements towards smaller sizes. To address this issue
we have estimated which fraction of our compact galaxies are potential
AGNs. Following Kauffmann et al. (2003) we have performed the BPT
diagram (Baldwin et al. 1981) log([OIII]/H$_{\beta}$)
vs. log([NII]/H$_{\alpha}$) to identify the AGNs in our
sample. Emission line fluxes of [OIII], H$_{\beta}$, [NII] and
H$_{\alpha}$ are measured by fitting small regions of the spectrum
with width $<$500\AA\, around the desired lines. The spectral region
is fitted using a linear combination of a gaussian (to model the
emission line) plus $\sim$40 stellar population models 
for the stellar absorption spectra. Based on the MILES stellar library
(S\'anchez-Bl\'azquez et al 2006; Cenarro et al 2007), these models
are an extension of those in Vazdekis (1999; hereafter V99$+$). Only
objects where the four emission lines are detected with S/N$>$3 are
considered. We use the demarcation to separate between AGNs and
starbursts proposed by Kauffmann et al. (2003; their Eq. 1). On doing
this we find only 2 galaxies in the compact sample that could be
considered as AGNs. Consequently the fraction of AGN galaxies in our
sample is very low ($\sim$7\%) and our results are not altered by
them.  Following the same criteria, we find that the number of AGNs
objects in the control sample is 29 (i.e. $\sim$10\%), similar to the
fraction found in the compact sample. This result reinforces the idea
that the presence of an AGN is not affecting our size estimates.

\section{Stellar Populations}

\begin{figure}

\plotone{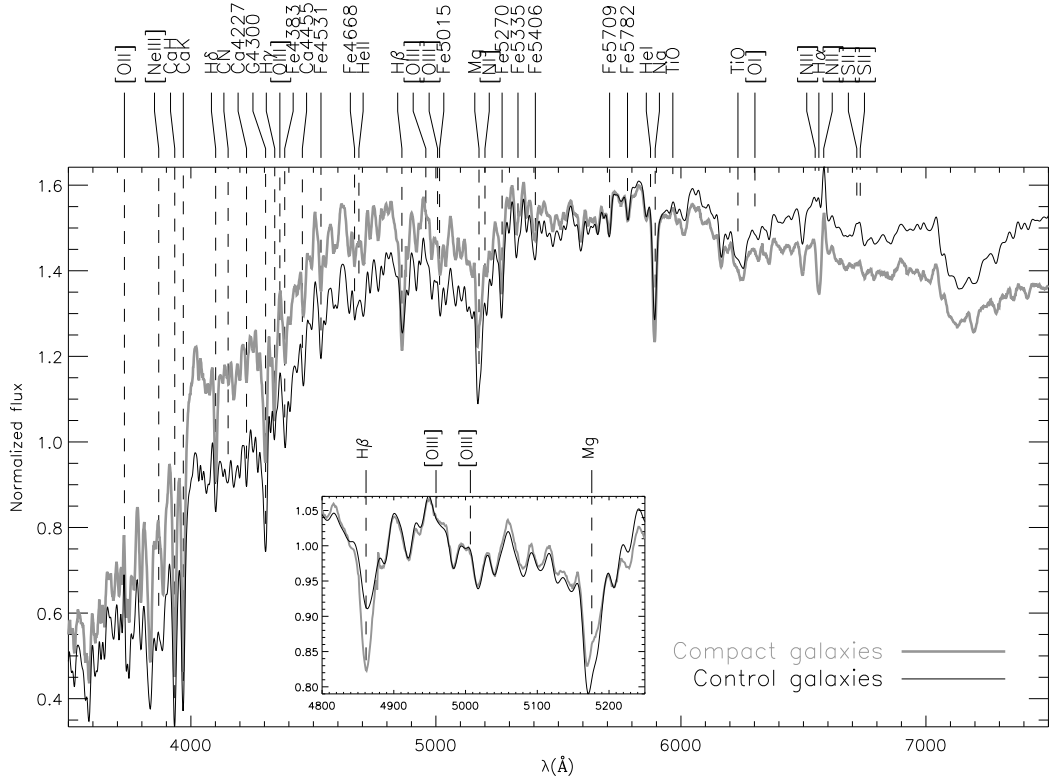}

\caption{Average mean spectra of the compact galaxies of our sample (grey thick line) and
the control galaxies (thin line). The individual spectra were previously convolved to the
highest dynamical velocity dispersion (320\,km\,s$^{-1}$) and divided by their
corresponding means, so they contribute with the same weight.  Each mean spectrum has been
normalized by its mean flux within the spectral range showed in each panel. The inset shows
the spectral regions around H$\beta$ and Mg lines.}

\label{spectra}

\end{figure}

To investigate whether the distinct structural properties of both
samples of galaxies are linked to differences in their stellar
population properties, we have analyzed both their global spectra
(Fig.~\ref{spectra}) and their luminosity-weighted ages and
metallicities on the basis of key absorption line-strength indices and
stellar population models.

\begin{figure}

\includegraphics[angle=-90,scale=0.7]{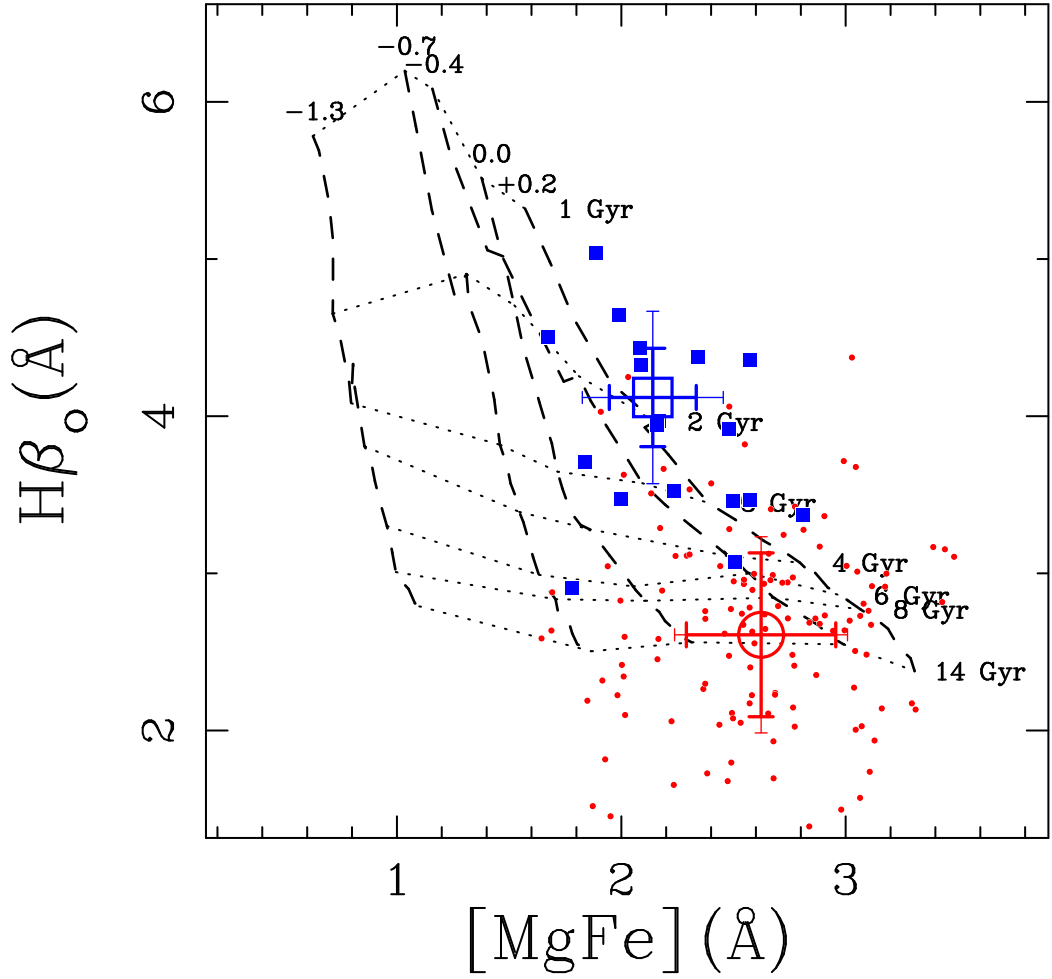}

\caption{H$_{\beta_0}$ vs [MgFe] diagram for 151 galaxies of the compact (filled squares,
18) and control samples (dots, 133) for which no $\lambda$5577\,\AA\ [O{\sc i}] skyline
residuals affect the index measurements. Overplotted are V99$+$ SSP model grids with
different ages (dotted lines) and metallicities (dashed lines) as given in the labels.  The
open square and circle are error-weighted mean indices for the compact and control samples
respectively. Thick error bars show the 1$\sigma$ index uncertainties, whereas the thin
error bars correspond to the dispersions of the samples.}

\label{indexes}

\end{figure}

Fig.~\ref{spectra} shows that the average compact galaxy looks clearly younger than the
average control galaxy. This is well seen from its bluer continuum and from the fact that
its H$_\beta$ absorption line is much stronger. Consequently, the metal lines of the
compact galaxies are weaker than those found in the galaxies of the control sample. A more
detailed disentangling of age and metallicity effects is presented in Fig. \ref{indexes}.
This figure shows the indices H$_{\beta_0}$ --an optimized age indicator by Cervantes \&
Vazdekis (2008)-- and [MgFe] --mainly sensitive to overall metallicity (Gonz\'alez 1993)--
measured for the compact (filled squares) and control galaxies (dots). All galaxies in the
range 0.132$\lesssim$z$\lesssim$0.159, with H$_{\beta_0}$ values subject to be affected by
strong $\lambda$5577\,\AA\ [O{\sc i}] skyline residuals, have been rejected from the figure
and subsequent analysis. To avoid systematics among the galaxy indices due to different
velocity dispersions, all the galaxy spectra were previously broadened to match the largest
$\sigma$ value of the whole sample, which is 320 km/s. Thus, taking into account the SDSS
instrumental broadening, all the indices in Fig.~\ref{indexes} are given at an overall
$\sigma$ of 332 km/s. Also, V99$+$ models are overplotted at the same spectral resolution
of the data.

Index uncertainties have been measured for each galaxy from the error
spectra provided in the SDSS database. The open square and circle in
Fig.~\ref{indexes} indicate, respectively, error-weighted mean indices
for the compact and control galaxy samples. For each of them, error
bars in thick lines illustrate the 1$\sigma$ typical index
uncertainties. In turn, thin error bars provide the error-weighted
r.m.s. standard deviations. The fact that the typical index errors are
smaller than their standard deviations seems to indicate that, within
each galaxy subsample, there exist certain differences --not explained
by errors-- among the stellar populations of their individual
galaxies. However, these differences are, by far, negligible as
compared to those between compact and control galaxies. As already
suggested by Fig.~\ref{spectra}, it is clear from Fig.~\ref{indexes}
that compact galaxies are much younger ($\sim 2$\,Gyr) than the
control galaxy sample, which is typically old ($\sim 14$\,Gyr). There
are also hints for compact galaxies having higher metallicities ([Z/H]
$\gtrsim +0.2$) than the control sample, with averaged values slightly
below solar.

It is worth noting here that the above age and metallicity results are difficult
to be driven by the larger apertures (a factor of 3 in units of r$_e$; see Section
2.3) sampled for the compact objects, as detailed studies of age and metallicity
gradients in normal ellipticals (e.g. S\'anchez-Bl\'azquez et al. 2006) report
that their outer parts are, on average, less metal rich and slightly older than
their centres. Therefore, if aperture issues were significantly affecting the
stellar population properties of the integrated spectra, we would expect compact
objects to be older and more metal poor. Also, note that the total galaxy
luminosity within 3r$_e$ is (for a S\'ersic surface brightness profile with a
n=4) only a 1.6 larger than that within 1r$_e$ (Trujillo et al. 2001), so if the
outer regions were the clue for the strong differences in luminosity-weighted age
between both galaxy subpopulations, compact objects should have unprecedented
young populations in their outskirts.

\section{Discussion}

The surprising young ages of the superdense massive galaxies in the nearby
universe cast some doubts about what are the mechanisms that most massive galaxies
in the high--redshift universe follow to reach their current sizes. At high
redshift the most massive galaxies are supposed to be the result of gas-rich
mergers resulting in a compact remnant (e.g. Khochfar \& Silk 2006; Naab et al.
2007). After this, dry mergers are expected to be the mechanism that moves these
very massive galaxies towards the current stellar mass size relation. Within this
scheme a non-negligible fraction (1-10\%) of superdense massive galaxies is
expected to survive intact since that epoch (Hopkins et al. 2008) and,
consequently, they are supposed to have old stellar populations which is at odds
with our findings. The above survival rate, however, is an upper limit  since only
major mergers (ratio 1:3) were considered at estimating this number density in
Hopkins et al. (2008). In fact, the role of minor mergers and small accretion can
be very relevant  (Naab et al. 2007)  at explaining the puffing up of galaxies.
Within the scheme of dry merging scenario, our result could  highlight the
importance of accounting for minor merging in order to make robust estimations in
the number density of old superdense massive galaxies in the present Universe.

If our compact galaxies are not relics of the
early universe how they could be form? A possibility is that they come from 
recent gas-rich disks merging. If this is the case we should not find many in
the present-day universe since the current massive disks population has not too
much gas. This could explain why the number of superdense young massive
galaxies is so scarce in the local Universe.

Recently, Fan et al. (2008) have suggested a puffing up scenario where the
superdense massive galaxies in the early universe can grow in size without
suffering merging. This mechanism, analog to the one proposed to explain the
growth of globular clusters (Hills 1980), argues that the observed strong
evolution in size is related to the quasar feedback, which removes huge amounts of
cold gas from the central regions quenching the star formation. If this mechanism
took place in all  massive galaxies in the past, no  old superdense massive galaxy
should be observed today as it seems to be the case. However, against this puffing
up scenario could be the stellar population ages of our galaxies. According to the
Fan et al. model, after the quenching of the star formation due to the gas
ejection, the galaxy needs some time to reach its new equilibrium configuration.
Following  their model, the amount of time is about 40 dynamical times. In massive
galaxies this will be around 2 Gyr. Interestingly, this is the typical mean
luminosity weighted age of the stellar population of our galaxies. However, we see
that they are still very compact. It is not clear how this can be fully
accomodated within the Fan et al. scenario.

A key element to explore the size evolution process is to the study the evolution
of the central velocity dispersion, $\sigma_\star$,  of the most massive galaxies
at a  given halo mass.  The two competing scenarios, "dry mergers" vs "puffing up"
predict a very different evolution of the central velocity dispersion  as cosmic
time evolves. In the merging scenario $\sigma_\star$ is basically constant with
time, at most a factor of 1.3 higher at high-z (Hopkins et al. 2008). In the
"puffing up" scenario the original (before the expansion of the object) central
velocity dispersion depends on their formation redshift as (1+z$_{form}$)$^{1/2}$.
As the object inflates, $\sigma_\star$ changes with the effective radius as
$\sigma_\star$$\propto$r$_e$$^{-1/2}$. Our results shows that galaxies in the
control (already "puffed up" in this scenario) and compact sample have very
similar $\sigma_\star$. If both type of objects were formed at high redshift our
results will be in contradiction with the "puffing up" scenario. If, on the other
hand, the nearby superdense massive galaxies have been formed recently, our
results agree with that scenario since young, low-z galaxies in the superdense
phase should have velocity dispersions similar to old "puffed up" galaxies,
because the effect of the increase of the radius essentially compensates the
effect of the different z$_{form}$.

To test  the prediction of the puffing up scenario  is  key to know whether the
nearby superdense massive galaxies are genuinally young objects. We need to
constrain whether our relatively young mean luminosity weighted age estimate of
$\sim$2 Gyr is representative of the whole stellar population of the objects or an
artifact due to a recent burst not involving a large amount of stellar mass. To
address this issue we have probed the Star Formation History (SFH) of our objects
by means of STARLIGHT (Cid Fernandes et al.  2005) exploring which SSP-model
combination best fits the observed spectra. Our preliminar results show that SFHs
from the compact galaxies are systematically different to those of the control
sample. For an average compact galaxy, more than 64\% of the luminosity comes from
stellar populations younger than 3 Gyr, in contrast to a 7\% in the control
sample. If this were the case, it will indicate that the superdense massive
galaxies are genuinally (structurally speaking) young objects. Finally, a
significant test to distinguish between  the above two competing scenarios will be
possible when central velocity dispersions of superdense massive galaxies at high
redshift will be determined.

\acknowledgments

IT and AJC acknowledge support from the Ram\'on y Cajal and Juan de la Cierva
Programs financed by the Spanish Goverment. We appreciate the contructive
comments of the referee that improved the quality of the manuscript. We thank
Ruym\'an Azzollini help on estimating the stellar masses and Jose Acosta and
Ana P\'erez for advice on AGN determination. Lulu Fan helped us on
understanding some relevant details of his size evolution model. This work has
been supported  by the Spanish Ministry of Education and Science grant 
\emph{AYA2007-67752-C03-01}.

\begin{table}
 \centering
 \begin{minipage}{140mm}

  \caption{Superdense massive galaxies sample}

  \begin{tabular}{ccccccc}
  \hline
ID NYU & Name &  R.A.    & Dec       &  M$_\star$           & r$_e$ & Redshift   \\
       &      &  (J2000) & (J2000)   &  (10$^{10}$M$_\sun$) & (kpc) &            \\
 \hline
       54829  & SDSS J153019.45-002918.6 &    232.58104  &    -0.4885095      &  8.39 & 1.12 &   0.085 \\
      155310  & SDSS J122705.10-031317.9 &    186.77130  &    -3.2216436      &  9.31 & 1.18 &   0.165 \\
      225402  & SDSS J113019.87+664928.8 &    172.58281  &     66.824716      &  9.67 & 1.30 &   0.143 \\
      265845  & SDSS J120032.46+032554.1 &    180.13528  &     3.4317179      &  8.01 & 1.31 &   0.143 \\
      321479  & SDSS J212052.74+110713.1 &    320.21978  &     11.120310      & 10.10 & 1.38 &   0.128 \\
      411130  & SDSS J082927.82+461331.4 &    127.36594  &     46.225414      &  8.86 & 1.30 &   0.167 \\
      415405  & SDSS J103050.53+625859.8 &    157.71053  &     62.983350      &  8.61 & 1.42 &   0.167 \\
      417973  & SDSS J090324.19+022645.3 &    135.85081  &     2.4459195      & 12.82 & 1.40 &   0.187 \\
      460843  & SDSS J143612.56+040411.8 &    219.05236  &     4.0699679      &  8.58 & 1.12 &   0.152 \\
      685469  & SDSS J222140.32+135914.2 &    335.41803  &     13.987279      &  9.03 & 1.38 &   0.149 \\
      721837  & SDSS J111136.18+534011.9 &    167.90070  &     53.670063      &  8.54 & 1.23 &   0.142 \\
      796740  & SDSS J144736.37+432945.7 &    221.90155  &     43.496021      &  9.76 & 1.47 &   0.182 \\
      807147  & SDSS J095705.52+045107.0 &    149.27301  &     4.8519610      &  8.78 & 1.47 &   0.162 \\
      815852  & SDSS J100629.34+071406.4 &    151.62225  &     7.2351152      &  9.35 & 1.49 &   0.121 \\
      824795  & SDSS J105324.14+062421.2 &    163.35062  &     6.4058952      & 10.26 & 1.49 &   0.186 \\
      890167  & SDSS J153934.07+441752.2 &    234.89197  &     44.297863      &  8.79 & 1.11 &   0.143 \\
      896687  & SDSS J143547.19+543528.7 &    218.94667  &     54.591381      &  9.23 & 1.43 &   0.130 \\
      929051  & SDSS J091926.44+065321.4 &    139.86017  &     6.8892928      &  8.25 & 1.44 &   0.184 \\
      986020  & SDSS J083917.44+303745.8 &    129.82268  &     30.629409      &  9.51 & 1.30 &   0.179 \\
     1044397  & SDSS J101637.23+390203.6 &    154.15510  &     39.034329      & 10.84 & 0.88 &   0.195 \\
     1173134  & SDSS J123238.80+425120.6 &    188.16165  &     42.855720      &  8.31 & 1.35 &   0.166 \\
     1689914  & SDSS J031406.38+001023.0 &    48.526604  &     0.1730749      &  8.95 & 1.35 &   0.163 \\
     1780650  & SDSS J120251.13+381644.2 &    180.71304  &     38.278960      &  8.23 & 1.47 &   0.157 \\
     1791371  & SDSS J120554.69+400958.9 &    181.47790  &     40.166362      &  9.14 & 1.06 &   0.154 \\
     1859261  & SDSS J091534.74+255606.2 &    138.89477  &     25.935065      &  8.10 & 1.24 &   0.155 \\
     2174994  & SDSS J235202.68+000244.2 &    358.01117  &     0.0456248      &  8.16 & 1.47 &   0.193 \\
     2258945  & SDSS J092723.34+215604.8 &    141.84726  &     21.934668      & 12.77 & 1.42 &   0.167 \\
     2402259  & SDSS J115032.32+170303.5 &    177.63473  &     17.050980      &  8.04 & 1.41 &   0.155 \\
     2434587  & SDSS J111659.35+170917.3 &    169.24737  &     17.154811      &  8.37 & 1.25 &   0.172 \\

\hline
\label{data}
\end{tabular}
\end{minipage}
\end{table}

\end{document}